\documentclass[10pt,twocolumn,twoside]{IEEEtran}

\usepackage{times}
\usepackage{amssymb}
\usepackage{latexsym}
\usepackage{amsmath}
\usepackage{epsfig}
\usepackage{amsfonts}
\usepackage{subfigure}

\newtheorem{algorithm}{\textit{Algorithm}}
\begin{document}

\title{Fast Algorithms for Sparse Recovery with Perturbed Dictionary}
\author{Xuebing~Han, Hao~Zhang, Gang Li
%\thanks{Submitted November, 2011; }% <-this % stops a space
\thanks{Xuebing Han is  with Guilin Air-Force Academy, Guilin, P. R. China (e-mail: thuhxb@gmail.com);}
\thanks{Hao~zhang and Gang Li are with the Department of Electronics Engineering, Tsinghua
University, Beijing, P. R. China (e-mail: haozhang@tsinghua.edu.cn, Gangli@tsinghua.edu.cn);}

%\thanks{This work was supported in part by the
%    National Basic Research Program of China (973 Program,
%    No.2010CB731901)}
 }

\maketitle

\begin{abstract}
In this paper, we account for approaches of sparse recovery from large underdetermined linear models with perturbation
present  in both the measurements and the dictionary matrix.  Existing methods have high computation and low efficiency.
The total least-squares (TLS) criterion has well-documented merits in solving linear regression problems while FOCal Underdetermined System Solver
(FOCUSS) has low-computation complexity in sparse recovery. Based on  TLS and FOCUSS methods, the present paper develops more fast and robust algorithms, TLS-FOCUSS and SD-FOCUSS. TLS-FOCUSS algorithm is not only
near-optimum but also fast in solving TLS optimization problems under sparsity constraints, and thus fit for large scale computation. In
order to reduce the  complexity of algorithm further, another
suboptimal algorithm named SD-FOCUSS is devised.  SD-FOCUSS can be applied in
MMV (multiple-measurement-vectors) TLS model, which fills the gap of solving linear regression problems under sparsity constraints. The convergence of TLS-FOCUSS algorithm and SD-FOCUSS algorithm is established with
mathematical proof.
The simulations illustrate the advantage of
TLS-FOCUSS and SD-FOCUSS in accuracy and stability, compared with  other
algorithms.
\end{abstract}


\section{Introduction}

The problem of finding sparse solutions to underdetermined system of linear equations has been a hot spot of researches in recent years, because of its widespread application in compressive sensing/sampling (CS)\cite{han17,han18}, biomagnetic imagining\cite{neuro}, source localization \cite{han13}, signal reconstruction\cite{han4,han16}, etc.

In the noise-free setup, CS theory holds promise to explain the equivalence between $\ell_0$-norm minimization and $\ell_1$-norm minimization as solving exactly linear equations when the unknown vector is sparse\cite{han3,restricted}. Variants of CS for "noise setup" of perturbed measurements are usually solved based on basis pursuit (BP) approach\cite{han1,han10} (utilizing method of linear programming\cite{han13} or Lasso\cite{han22}), greedy algorithms (e.g. OMP\cite{han15}, ROMP\cite{ROMP}, CoSaMP\cite{han14}, etc) or least-squares methods with $\ell_1$-regularization (e.g., FOCUSS\cite{han4,han5,han16}). However, exiting BP, greedy algorithms and FOCUSS do not account for perturbations present in the dictionary matrix, i.e. regression matrix.

Recently, only a little attention has been paid on the sparse problems with perturbations present both in measurements and dictionary matrix. Performance analysis of CS and BP methods for the linear regression model under sparsity constraints was researched in \cite{han7}, \cite{effect_basis_mismatch} and \cite{sensitivity_basis_mismatch}; a feasible approach in \cite{han8} , named S-TLS, was devised to reconstruct sparse vectors based on Lasso from the "fully-perturbed" linear model. However, the research of \cite{han7}, \cite{effect_basis_mismatch} and \cite{sensitivity_basis_mismatch} are limited in theoretical aspect and do not devise systematic approaches. Due to its highly-computational burden, S-TLS is very time-consuming, and thus unsuitable for large scale problems.

In this paper, an extension form of FOCUSS is devised solving sparse problems to "fully-perturbed" linear model. Belonging to categories of convex optimization, LP and Lasso  have the stable results but their computational burden is the highest; greedy algorithms have low computation, but their performances can only be
guaranteed when the dictionary matrix satisfies some
rigorous conditions, such as very small restricted isometry
constants \cite{han3}. FOCUSS was originally designed to obtain a
sparse solution by successively solving quadratic optimization
problems  and was widely used to deal with compressed sensing problems. The obvious advantages of FOCUSS are its low computation and stable results.
For FOCUSS, only a few iterations tends to be enough to obtain a
rather good approximating solution.
%Experimental results in \cite{han16} proved that FOCUSS performed better than greedy algorithm (OMP) under MMV model.
So it is an excellent choice to develop FOCUSS to solve approximate sparse solutions to linear regression model, especially  in large scale application.

Our objective is to overcome  the
influence of perturbation present in dictionary matrix and measurements on the accuracy of sparse recovery effectively. Meanwhile, the merits of FOCUSS, rapid convergence
and good adaption to intrinsic properties of dictionary matrix, are
maintained. First, objective function to be optimized can be
obtained under a Bayesian framework. Then the necessary condition
for the optimizing solution is that each first-order partial
derivative of objective function is equal to zero. Next we can
get the iterative expression using  iterative relaxation algorithm.
Finally, the new algorithms are proved to be convergent.

The paper is organized as follows. In Section \ref{sec1}, we
introduce perturbed linear regression model for sparse recovery,
and analyze the optimal problem simply. In Section \ref{sec2}, we
use a MAP estimate to obtain the objective function to be optimized,
then yield an iterative  algorithm to provide solutions, named
TLS-FOCUSS for adopting TLS method and framework of FOCUSS.
Convergence of TLS-FOCUSS is proved. In Section \ref{sec3}, we
propose another  algorithm based on FOCUSS and TLS model, named
SD-FOCUSS to distinguish TLS-FOCUSS. Though SD-FOCUSS is a
suboptimal optimal, its computation is low and it can be used in MMV
case.
%In Section \ref{sec4}, we analyze the complexity of
%TLS-FOCUSS, SD-FOCUSS and S-TLS.
In the simulation of Section
\ref{sec5}, the performances of mentioned algorithms are presented.
Finally, we draw some conclusions  in Section \ref{sec6}.

\section{Perturbed Linear Regression Model}\label{sec1}

Consider the underdetermined  linear system of $\textbf{y}=A\textbf{x}$, where $A$
is an $m\times n$ matrix with $m<n$, $\textbf{y}$ is the given $m\times 1$ data vector,
and $\textbf{x}$ is unknown $n\times 1$ vector to be recovered. With $\textbf{x}$ being
sparse, and $A$ satisfying some property (e.g., RIP\cite{han3}), CS theory asserts that
exact recovery of $\textbf{x}$ can be guaranteed by solving the convex problem\cite{han1,han3,RIP}: $\min_x\|\textbf{x}\|_1
~~\textrm{s.t.} (\textrm{subject~to})~\textbf{y}=A\textbf{x}$. Suppose that data perturbations exist in the linear model $A\textbf{x}$. The corresponding convex problem can be written as a Lagrangian form\cite{han1,han13,han5}: $
\min_\textbf{x}\|\textbf{y}-A\textbf{x}\|_2^2+\gamma\|\textbf{x}_p^p$, where $\|\cdot\|_p^p=\sum|\cdot|^p$, $\gamma>0$ is a sparsity-tuning parameter\cite{han8}, and $0<p\leq 1$ ($p$ is set to 1 in \cite{han1,han8}). What the present paper focusses on is how to reconstruct sparse vector efficiently from over- and especially under-determined linear regression models while perturbations are present in $\textbf{y}$ and/or $A$.

  The perturbed linear
regression model can be formulated as follows\cite{han9,han12}:
\begin{equation}\label{eq-1}
\textbf{y}=(A+E)\textbf{x}+\textbf{e},
\end{equation}
where $\textbf{e}$ represents perturbation vector and $E$ represents perturbation matrix. Due to randomness and uncertainty, it is usually assumed that the components of noise in the same channel are independently and identically Gaussian distributed, e.g. $\textbf{e}\sim N(0,\sigma_1^2I)$ and $ \textrm{vec}(E)\sim N(0,\sigma_2^2I)$, where vec$(\cdot)$ is matrix vectorizing operator.

(\ref{eq-1}) can be rewritten as
\begin{equation*}
(B+D)\left[ {\begin{array}{*{20}{c}}
{1}\\
{\textbf{x}}
\end{array}} \right]=0,
\end{equation*}
where $B=[-\textbf{y},A],{\quad}D=[\textbf{e},E]$. Without exploiting sparsity, TLS has well documented merits solving above problem. For over-determined models TLS estimates are given by
\begin{equation*}%%mark
  \hat{\textbf{x}}=\arg\min_{D,\textbf{x}}\|D\|_F^2,\quad \textrm{s.t.}(B+D)\left[
  \begin{array}{c}
    1\\
    \textbf{x}
  \end{array}
  \right]=0,
\end{equation*}
where $\|\cdot\|_F^2$ represents Forbenius-form operator. With the assumption of $\textrm{vec}(D)\sim N(0,\sigma^2)$,  \cite{han12} gives the equivalent solutions as
\begin{equation}
  \hat{\textbf{x}}=\arg \min_\textbf{x} \frac{\|\textbf{y}-A\textbf{x}\|^2}{1+\|\textbf{x}\|^2}
\end{equation}

The distinct objective of the present paper is twofold: developing efficient solvers for fully-perturbed linear models, and accounting for sparsity of $\textbf{x}$. To achieve these goals, following optimization problem must be sovled
\begin{equation}\label{eq-13}
\hat{\textbf{x}}=\arg\min_{\textbf{x},D}\left[\|D\|_F^2+\gamma\|\textbf{x}\|_p^p\right],
\end{equation}
where $\gamma >0$, and $0<p\leq1$. In (\ref{eq-13}), the $\ell_F$-term forces the quadratic sum of
perturbations to be minimal while the $\ell_p$-term forces sparsity
of recovery\cite{han1,han8}, and $\gamma$ controls tradeoff
between above two terms. Developing  efficient algorithms to get the local even global optimum of (\ref{eq-13}) is the main goal. In next section, we will explain how to get
the objective function  and estimate the value of
$\gamma$ with a bayesian formation, then develop the new method of
optimization.

\section{TLS-FOCUSS Algorithm}\label{sec2}

This section develops an extension of FOCUSS, TLS-FOCUSS, to solve
(\ref{eq-1}) using Bayesian framework \cite{han1} and main idea of
TLS. For simplifying formulas, we assume $\sigma=\sigma_1=\sigma_2$, that is $\textrm{vec}(D)\sim N(0,\sigma^2)$. At the end of the section, we will introduce how to process the situation with $\sigma_1\neq \sigma_2$.

\subsection{Bayesian Formulation}\label{sec3-1}

From (\ref{eq-1}), we obtain
\begin{equation}\label{eq-14}
\textbf{y}-A\textbf{x}=G(\textbf{x})\textbf{v},
\end{equation}
where $G(\textbf{x})=[1,\textbf{x}^T]\otimes{I_{m\times{m}}}$,
$\textbf{v}=\textrm{vec}(D)$, ($\otimes$ represents Kronecker product).
Under Bayesian viewpoint, unknown vector $\textbf{x}$ is assumed to be
random and independent of $D$. Then the MAP
estimation of $\textbf{x}$ can be obtained as:
\begin{align}\label{map}
\hat{\textbf{x}}_{\textrm{MAP}}&=\arg\max_\textbf{x}\ln{p(\textbf{x}|\textbf{y})} \nonumber\\
&=\arg\max_\textbf{x}[\ln{p(\textbf{y}|\textbf{x})}+\ln{p(\textbf{x})}].
\end{align}
This formula is general and offers considerable flexibility. In order to obtain optimality of the resultant estimates, another assumption must be made on the distributions of the  solution vector $\textbf{x}$. As discussed in \cite{han5}, the elements of sparse $\textbf{x}$ are assumed to be distributed as general Gaussian and independent,
\begin{equation}\label{eq-3}
p(\textbf{x})=C_2\exp\left(-\frac{1}{2\beta^p}\sum_{k=1}^m\big|\textbf{x}[k]\big|^p\right),
\end{equation}
where $C_2$ is constant, $0<p\leq1$ and $\beta$ is
constant depended on $p$ with $\beta=2^{-\frac{p}{2}}\frac{\Gamma(1/p)}{\Gamma(3/p)}$ (
where $\Gamma(\cdot)$ means Gamma function). Only one parameter characterizes the distribution in (\ref{eq-3}). The pdf moves toward  a uniform distribution as $p\to \infty$ and toward a very peaky distribution as $p\to0$.

With $\textbf{v}{\sim}N(0,\sigma^2I)$ and
$G(\textbf{x})G^H(\textbf{x})=(1+\|\textbf{x}\|^2)I$, we have
\begin{equation}\label{eq-2}
\ln{p(\textbf{y}|\textbf{x})}=\frac{1}{2\sigma^2}\frac{(\textbf{y}-A\textbf{x})^H(\textbf{y}-A\textbf{x})}{1+\|\textbf{x}\|^2}+C_1,
\end{equation}
where $C_1$ is constant. With the densities of the perturbation vector $\textbf{v}$ and the solution vector $\textbf{x}$, we can now proceed to find the MAP estimate as
\begin{equation}\label{eq-4}
\hat{\textbf{x}}_{\textrm{MAP}}=\arg\min_\textbf{x}\left[\frac{\|\textbf{y}-A\textbf{x}\|_2^2}{1+\|\textbf{x}\|_2^2}+\gamma\|\textbf{x}\|_p^p\right],
\end{equation}
where $\gamma=\sigma^2/\beta^p$.

%Actually, how to choose  parameter $\gamma$ is not limited to the
%method this paper mentioned. One can find the other methods from
%\cite{han5} and \cite{han23}.

\subsection{Derivation of TLS-FOCUSS}

The optimization problem (\ref{eq-4}) is equivalent to
\begin{align}\label{eq-12}
\begin{array}{c}
\arg\min_{\textbf{z}'}J(\textbf{z}')\\
\textrm{where}\quad
J(\textbf{z}')=\left[\frac{\|B\textbf{z}'\|_2^2}{\|\textbf{z}'\|_2^2}+\gamma\|\textbf{z}'\|_p^p\right],
\end{array}
\end{align}
with
\begin{equation}\label{eq-15}
\textbf{z}'=\left[\begin{array}{c} 1\\
\textbf{x}
\end{array}\right],{\quad}B=[-\textbf{y},A].
\end{equation}
To simplify the objective function,we normalize  $\textbf{z}'$ and get the equivalent form as
\begin{equation}
\min_\textbf{z}\left[\left\|B\textbf{z}\right\|_2^2+\gamma\left\|\textbf{z}\right\|_p^p\right],\quad\textrm{s.t.}\
\|\textbf{z}\|_2^2=1.
\end{equation}
Using Lagrange multiplier method, the objective function can be
rewritten as
\begin{equation}\label{eq-temp}
T(\textbf{z})=\|B\textbf{z}\|_2^2+\gamma\|\textbf{z}\|_p^p+\lambda(1-\textbf{z}^H\textbf{z}),
\end{equation}
where $\lambda$ is the Lagrange multiplier. The factored gradient approach developed in \cite{affine}, an iterative method can be derived to minimize $T(\textbf{z}$. A necessary condition for the optimum solution $\textbf{z}_*$ is that it must satisfy $\nabla_zT(\textbf{z}_*)=0$. We can get
\begin{align}\label{eq-5}
(B^HB+\alpha\Pi(\textbf{z}_*))\textbf{z}_*&=\lambda\textbf{z}_*,
\end{align}
where
\begin{equation*}
\alpha=p\gamma/2,{\quad}
\Pi(\textbf{z})=\textrm{diag}\left(\left[\left|\textbf{z}[i]\right|^{p-2}\right]_{i=1,\cdots,n+1}\right).
\end{equation*}
So the iterative relaxation scheme can be constructed as
\begin{equation}\label{eq-6}
\big(B^HB+\alpha\Pi(\textbf{z}_{k-1})\big)\textbf{z}_k=\lambda\textbf{z}_k.
\end{equation}
It is easily seen that $\lambda$ should be the minimal eigenvalue of
objective matrix $B^HB+\alpha\Pi(\textbf{z}_{k-1})$. However, it's very hard to
find it for two reasons: firstly, the minimal eigenvalue is likely
close to zero because objective matrix  is approximately singular;
secondly, the dimension of matrix above is tremendous for most large
scale application, which leads to a big computational burden for
matrix inversion.  (\ref{eq-6}) implies that
\begin{equation}\label{eq-7}
\big(B^HB+\alpha\Pi(\textbf{z}_{k-1})\big)^{-1}\textbf{z}_k=\frac{1}{\lambda}\textbf{z}_k.
\end{equation}
From (\ref{eq-7}), finding the minimal eigenvalue is taken place of by
finding the maximal eigenvalue. The latter become much more
well-posed. Moreover, with the aid of matrix inversion formula, we
have
\begin{align}
&\big(B^HB+\alpha\Pi(\textbf{z}_{k-1})\big)^{-1}\nonumber\\
=&\frac{1}{\alpha}\Big(W_k^2-W_k^2B^H\big(\alpha{I}-BW_k^2B^H\big)^{-1}BW_k^2\Big),
\end{align}
where $W_k^2=\Pi^{-1}(\textbf{z}_{k-1})$. Let
\begin{equation}
\Phi_k=W_k^2-W_k^2B^H(\alpha{I}-BW_k^2B^H)^{-1}BW_k^2,
\end{equation}
then we obtain
\begin{equation}\label{eq-8}
\Phi_k\textbf{z}_k=\frac{\alpha}{\lambda}\textbf{z}_k.
\end{equation}

It should be mentioned that the dimension of matrix
\mbox{$\alpha{I}-BW_k^2B^H$} is much less than that of matrix
$B^HB+\alpha\Pi(\textbf{z}_{k-1})$, so the cost of matrix inversion
is extremely reduced. Besides, we need only calculate the
maximal eigenvalue and corresponding eigenvector instead of all the
eigenvalue and eigenvector of $\Phi_k$. That is to say,
some highly efficient solver, such as Lanczos iteration, could be
utilized to make the problem further simplified.

Noting that the optimal problem (\ref{eq-4}) is not global convex,  the
TLS-FOCUSS algorithm guarantees convergence to a local optimum. Once
the initial point $\textbf{z}_0$ is close to the true point,
estimation of true value can be found through iterations. In this
paper, we set $\textbf{x}_0=A^H(AA^H)^{-1}\textbf{y}$, then
$\textbf{z}_0$ is set through substituting $\textbf{x}_0$ into
(\ref{eq-15}) and normalization of $\textbf{z}_0'$.

When  the convergent solution $\textbf{z}_*$ is obtained, we can get
\begin{equation}\
\textbf{x}_{TLS-FOCUSS}=[z_{*2},\cdots,z_{*{n+1}}]^T/z_{*1}.
\end{equation}

\textbf{Algorithm 1} is the algorithmic description of TLS-FOCUSS.

\begin{algorithm}[TLS-FOCUSS]
%\caption{TLS-FOCUSS}
\mbox{}
\begin{itemize}
\item[] \textbf{Input:} $\textbf{z}_0$, $B$, $\alpha$, $p$.

\item[1] Set
$W_k=\textrm{diag}\left(\left[\left|\textbf{z}_{k-1}[i]\right|^{1-\frac{p}{2}}\right]_{i=1,\cdots,n+1}\right)($,
and $p\in(0,1]$);

\item[2] Calculate $\Phi_k=W_k^2-W_k^2B^H(\alpha{I}+BW_k^2B^H)^{-1}BW_k^2$.

\item[3] Compute the largest eigenvalue $\lambda_k$ and corresponding eigenvector $\textbf{u}_k$ of $\Phi_k$ using  Lanczos method.

\item[4] Set $\textbf{z}_k=\textbf{u}_k$.

\item[5] If $\|\textbf{z}_k-\textbf{z}_{k-1}\|_2^2/\|\textbf{z}_{k-1}\|_2^2<\epsilon$, exit; else goto step 1.
\end{itemize}
\end{algorithm}

\subsection{Convergence and Sparsity}

To show that TLS-FOCUSS algorithm can approximately solve  the sparse problem of (\ref{eq-1})
 through iterative method, two key results should be obtained:
 i) TLS-FOCUSS is a convergent algorithm that it indeed reduces $J(\textbf{z})$ at each iterate step; ii) the convergence points of TLS-FOCUSS are sparse.

\begin{proof}[proof of convergence]
  From (\ref{eq-6}) we have
\begin{equation}
B_W^HB_W\textbf{q}_k+\alpha \textbf{q}_k-\lambda
W_k^2\textbf{q}_k=0,
\end{equation}
where $B_W=BW_{k}$, $\textbf{q}_k=W^{-1}_{k}\textbf{z}_k$. And
$\textbf{q}_k$ can be treated as an optimal solution:
\begin{equation}\label{eqq}
\textbf{q}_k=\arg\min_\textbf{q}\left[\|B_W\textbf{q}\|^2+\alpha\|\textbf{q}\|^2+\lambda
(1-\textbf{q}^HW_{k}^2\textbf{q})\right].
\end{equation}
From (\ref{eqq}) and the equivalence of optimization between  (\ref{eq-12}) and (\ref{eq-temp}),
$\textbf{z}_k$ can be expressed a solution to an optimization
problem:
\begin{align}\label{eq-11}
\textbf{z}_k=&\arg\min Q_k(\textbf{z}),\nonumber\\
\textrm{where} ~
Q_k(\textbf{z})=&\frac{\|B\textbf{z}\|^2}{\|\textbf{z}\|^2}+\alpha
\|W_k^{-1}\textbf{z}\|^2.
\end{align}
So  TLS-FOCUSS algorithm can be considered
to be a method of re-weighted $\ell_2$-form minimization  \cite{han4,han5}. Since $\textbf{z}_k$ is the local
unique solution to minimize $Q_k(\textbf{z})$, we have
\begin{equation}\label{eq-20}
Q_{k}(\textbf{z}_k)<Q_{k}(\textbf{z}_{k-1})
\end{equation}
with $\textbf{z}_k,~\textbf{z}_{k-1}$  located in the same small domain and $\textbf{z}_k\neq \textbf{z}_{k-1}$.

  And we can get the conclusion \cite{han5}
that
\begin{align}\label{eq27}
&\sum_i \left(|{z}_2[i]|^p-|{z}_1[i]|^p\right)\nonumber\\
    \leq &\sum_i \frac{p}{2}|z_1[i]|^{p-2}\left(z_2[i]^2-z_1[i]^2\right)\nonumber\\
=&\frac{p}{2}\left[\textbf{z}_2^T\Pi(\textbf{z}_1)\textbf{z}_2-\textbf{z}_1^T\Pi(\textbf{z}_1)\textbf{z}_1\right],
\end{align}
where $\Pi(\textbf{z})=diag(|z[i]|^{p-2})$. With $\textbf{z}_{k-1}$
and $\textbf{z}_k$ ($\textbf{z}_k\neq \textbf{z}_{k-1}$) obtained from
the $(k-1)$th and $k$th iteration of TLS-FOCUSS,
we have
\begin{align}\label{eq-21}
&J(\textbf{z}_k)-J(\textbf{z}_{k-1})\nonumber\\
\leq &\left[{\frac{\|B\mathbf{z}_k\|_2^2}{\|\mathbf{z}_k\|_2^2}+
{\alpha\mathbf{z}_k^TW_k^{-2}\mathbf{z}_k}}\right]-
\left[{\frac{\|B\textbf{z}_{k-1}\|_2^2}{\|\textbf{z}_{k-1}\|_2^2}+
{\alpha\mathbf{z}_{k-1}^TW_k^{-2}\mathbf{z}_{k-1}}}\right]\nonumber\\
=&Q_k(\textbf{z}_k)-Q_k(\textbf{z}_{k-1})<0,
\end{align}
where $\textbf{z}_{k}$ and $\textbf{z}_{k-1}$ are obtained from the $k\textrm{-th}$ and ($k-1$)-th iteration step of TLS-FOCUSS. The first inequality follows from (\ref{eq27}) and the last inequality from (\ref{eq-20}). So the value of $J(\textbf{z}_k)$ decreases as $k$ increases. From (\ref{eq-21}) and $J(\textbf{z}_k)\geq 0$, it can be concluded that  TLS-FOCUSS is a convergent algorithm.
\end{proof}

\begin{proof}[proof of sparsity]
Assuming $\textbf{z}_0$ is a local minima of $J(\textbf{z})$, $\textbf{z}_0$ is also a local minima to an optimization problem: $  \min\limits_\textbf{z}\sum\limits_{i} |z[i]|^p\quad
    \textrm{s.t.}~ (B+D)\textbf{z}=0$,
which can be rewritten as
\begin{equation}\label{p-norm}
  \min_\textbf{x}\sum_i\left|x[i]\right|^p\quad\textrm{s.t.}~\textbf{y}=(A+E)\textbf{x}+\textbf{e}.
\end{equation}
Similarly shown in \cite{han13,han5,stable} (especially $p=1$), as an equivalence of $\ell_0$-norm optimization
 above optimization problem can obtain the local minima which are necessary sparse. The provement of equivalence between $\ell_0$-norm and $\ell_p$-norm about fully-perturbed model is aslo an open problem.

 Let $\textbf{z}_*$ be an fixed point of the algorithm, and therefore a solution of (\ref{eq-7}). If $\textbf{z}_*$ is not sparse, it is not a local minima of (\ref{p-norm}), so there must be other points close to $\textbf{z}_*$ which can reduce $J(\textbf{z})$\cite{affine}. Thus it can be concluded that only sparse solutions are stable points of TLS-FOCUSS algorithm.
\end{proof}

\subsection{Robust Modification}

Note that we  assumed the components of perturbation matrix
$[\textbf{e},E]$ are i.i.d.
(independent and identically distributed). Actually, only  noise existing in the same channel is assumed to be i.i.d..
 When $\textbf{e}$ and $E$ have the different distributed variances,  it is necessary to
normalize variances of perturbations before signal reconstruction. Assume that
$\textbf{e}$ and $E$ are independent, and $\textbf{e}\sim
N(0,\sigma^2I_1)$, $\textrm{vec}(E)\sim N(0,\sigma_2^2I_2)$. Then
 we have $\textbf{y}-A\textbf{x}=G(\textbf{x})\textbf{v}$ with
\begin{equation*}
G(\textbf{x})=\left[1,\frac{\sigma_2}{\sigma}\textbf{x}^T\right]\otimes{I_{m\times{m}}},~
\textbf{v}=\left[\begin{array}{c} \textbf{e}\\
\frac{\sigma}{\sigma_2}\textrm{vec}(E).
\end{array}\right]
\end{equation*}
It can be seen $\textbf{v}\sim N(0,\sigma^2I)$. For
(\ref{eq-12}), instead of (\ref{eq-15}) we have
\begin{equation*}
  \textbf{z}'=[1,\frac{\sigma_2}{\sigma}\textbf{x}^T]^T,{\quad}B=[-\textbf{y},\frac{\sigma}{\sigma_2}A].
\end{equation*}
Now TLS-FOCUSS algorithm can be used to recover the sparse signal.

\section{SD-FOCUSS Algorithm}\label{sec3}

TLS-FOCUSS needs to compute the maximal eigenvalue  and its
corresponding eigenvector of matrix $\Phi_k$ in every iteration. By utilizing
Lanczos algorithm,  TLS-FOCUSS algorithm can be speeded up greatly.
However, it is still possible to release much
more the computation burden  while the performance descends a little. In this section, a
suboptimal algorithm, named SD-FOCUSS (Synchronous Descending
FOCUSS), is divised.

Based on TLS model (\ref{eq-1}),  Zhu in \cite{han8} devised a
sparse recovery algorithm S-TLS. To optimize the objective function,
 S-TLS adopted iterative block coordinate  descent
method, yielding successive estimates of $\textbf{x}$ with $E$ fixed
and alternately of $E$ with $\textbf{x}$ fixed until obtaining
stable solutions. The algorithm needs several convergent procedures
before final convergence. Different from S-TLS, SD-FOCUSS is more
efficient, which only needs one convergent procedure, with estimating
$\textbf{x}$ and $E$ synchronously in each iteration; meanwhile, SD-FOCUSS has lower computation complexity.

\subsection{Bayesian Formulation}

In this section, $\textbf{x}$ and $E$ in (\ref{eq-1}) are both considered
variants to be optimized.
Assume that $e\sim N(0,\sigma^2I_1)$,
$\textrm{vec}(E)\sim N(0,\sigma_2^2I_2)$, and $e$, $E$ are
independent. So we have
\begin{align}
  p_\textbf{e}(\textbf{e})&=C_3\exp\left(-\frac{\textbf{e}^H\textbf{e}}{2\sigma_1^2}\right)\nonumber\\
  p_{E}({E})&=C_4\exp\left(-\frac{\textrm{vec}({E})^H\textrm{vec}({E})}{2\sigma_2^2}\right)=\exp\left(-\frac{\|{E}\|_{F}^2}{2\sigma_2^2}+C_2\right)
\end{align}
Where $C_1$, $C_2$ are constant. The Bayesian formulation is described as
\begin{align}\label{eq-18}
  &[\hat{\textbf{x}}_\textrm{MAP},\hat{E}_\textrm{MAP}]=\arg\max_{\textbf{x},E}\ln
  p(\textbf{x},E|\textbf{y})\nonumber\\
  =&\arg\max_{\textbf{x},E}\left[\ln p(\textbf{y}|\textbf{x},E)+\ln
  p(\textbf{x})+\ln p(E)\right].
\end{align}
Here we have
\begin{align}\label{eq-19}
  \ln p(\textbf{y}|\textbf{x},E)=\frac{1}{\sigma_1^2}\left\|\textbf{y}-\left(A+E\right)\textbf{x}\right\|_2^2+\ln C_3.
\end{align}

\subsection{Derivation of SD-FOCUSS}
From (\ref{eq-3}) (\ref{eq-18}) and (\ref{eq-19}), the objective
function can be written as
\begin{equation}
J(\textbf{x},E)=\left\|\textbf{y}-\left(A+E\right)\textbf{x}\right\|_2^2+
\frac{\sigma^2}{\sigma_2^2}\textrm{tr}(E^HE)+\gamma\|\textbf{x}\|_p^p
\end{equation}
where $\textrm{tr}(\cdot)$ means trace of matrix and
$\textrm{tr}(E^HE)=\|E\|_F^2$. The necessary condition of the
optimal solution satisfies that partial differentiation to each
component for $J(\textbf{x},E)$ is equal to zero, that is:

a)  $\nabla_EJ(x,E_*)=0$.

We can get
\begin{equation*}
  \nabla_EJ(\textbf{x},E)=E{\textbf{x}}\textbf{x}^H-(\textbf{y}-A\textbf{x})\textbf{x}^H
  +\sigma_1^2\sigma_2^{-1}E.
\end{equation*}
So we can get the estimate of $E$ as a function of $\textbf{x}$:
\begin{equation}\label{eq6}
  {E_*(\textbf{x})}=\frac{(\textbf{y}-A\textbf{x})\textbf{x}^H}
  {\sigma_1^2\sigma_2^{-2}+\textbf{x}^H\textbf{x}}.
\end{equation}
Here the fact of $(\lambda I+F^HF)^{-1}F^H=F^H(\lambda
I+FF^H)^{-1}$ is used.

b) $\nabla_xJ(x_*,E)=0$.

Referring to
\cite{han5}, we can get the iterative relaxation scheme of
$\textbf{x}$ as
\begin{equation}\label{eq-16}
\textbf{x}_k=W_kA_k^H(A_kA_k^H+\alpha I)^{-1}\textbf{y},
\end{equation}
where $\alpha=\frac{p\gamma}{2}$,
$W_k=\textrm{diag}\left(\left[|\textbf{x}_{k-1}[i]|^{1-\frac{p}{2}}\right]_{i=1,\cdots,n}\right)$
 and $A_k=\big(A+E(\textbf{x}_{k-1})\big)W_k$. There exists error
inevitably when we estimate $E$, thus accuracy of estimating
$\textbf{x}$ will be affected. It is a suboptimal algorithm.

\textbf{Algorithm 2} is the algorithmic description of SD-FOCUSS.

\begin{algorithm}[SD-FOCUSS]
%\caption{SD-FOCUSS}
\mbox{}
\begin{itemize}
\item[] \textbf{Input:} $y$, $\textbf{x}_0$, $E_0$ $A$, $\sigma$, $\sigma_2$, $p$.

\item[1] Set
$W_k=\textrm{diag}\left(\left[\left|\textbf{x}_{k-1}[i]\right|^{1-\frac{p}{2}}\right]_{i=1,\cdots,n}\right)($,
and $p\in(0,1])$;

\item[2] Calculate
\[{E_k} = \frac{{({\bf{y}} - A{{\bf{x}}_{k - 1}}){\bf{x}}_{k - 1}^H}}{{{\sigma ^2}\sigma _2^{ - 2} + \left\| {{{\bf{x}}_{k - 1}}} \right\|_2^2}},~\textrm{and}~A_k=(A+E_{k})W_k;\]
%$E_k=\frac{(\textbf{y}-A\textbf{x}_{k-1})\textbf{x}_{k-1}^H}{\sigma^2\sigma_2^{-2}+\left\|\textbf{x}_{k-1}\right\|_2^2}$,
 %and $$;

\item[3] Calculate $\textbf{x}_k=W_kA_k^H(A_kA_k^H+\alpha
I)^{-1}\textbf{y}$;

\item[4] If $\|\textbf{x}_k-\textbf{x}_{k-1}\|_2^2/\|\textbf{x}_{k-1}\|_2^2<\epsilon$, exit; else goto step 1.
\end{itemize}
\end{algorithm}

\subsection{Proof of Convergence}

Formula (\ref{eq-16}) can be seen as $\textbf{x}_k=W_k\textbf{b}_k$,
where $\textbf{b}_k$ can be treated as an optimal solution, that is
\begin{equation}
  \textbf{b}_k=\arg\min
  \|\textbf{y}-A_kW_k\textbf{b}\|_2^2+\alpha\|\textbf{b}\|_2^2
\end{equation}
Alternately and equivalently, $\textbf{x}_k$ can be expressed a
solution to an optimization problem:
\begin{align}
  \textbf{x}_k&=\arg\min_xQ_k(\textbf{x}),\nonumber\\
  \textrm{where}\quad Q_k(\textbf{x})&=\|\textbf{y}-A_k\textbf{x}\|_2^2+\alpha\|W_k^{-1}\textbf{x}\|_2^2.
\end{align}
Referring to (\ref{eq-20})-(\ref{eq-21}), we can conclude  that
SD-FOCUSS is also a convergent algorithm.

\subsection{SD-FOCUSS Extension: MMV case}

Besides low computation, the breakthrough advantage of SD-FOCUSS is
that it can be used in multiple measurement vectors (MMV) model,
while TLS-FOCUSS and S-TLS \cite{han8} cannot fit this model or
remain to be developed. Supposed
$\textbf{y}^{(l)}=(A+E)\textbf{x}^{(l)}+\textbf{e}^{(l)}$, with
$l=1,\cdots,L$, where $\textbf{y}^{(l)}\in R^m$ and
$\textbf{x}^{(l)}\in R^n$. Suppose that the vectors
$\textbf{x}^{(l)},l=1,\cdots,L$ are sparse and have the \emph{same}
sparsity profile, and let
$Y=[\textbf{y}^{(1)},\cdots,\textbf{y}^{(L)}]$,
$X=[\textbf{x}^{(1)},\cdots,\textbf{x}^{(L)}]$.

The objective function for MMV case is expressed as
\begin{align}
J(X,E)=&\left\|Y-\left(A+E\right)X\right\|_F^2+\nonumber\\
&\frac{\sigma^2}{\sigma_2^2}\|E\|_F^2+\gamma\sum_{i=1}^n\left(\sum_{l=1}^Lx^{(l)}[i]^2\right)^{p/2}
\end{align}
The weight matrix $W_k$ can be re-expressed as \cite{han16}
\begin{align}
  W_k=&\textrm{diag}\left(c_k[i]^{1-p/2}\right)\quad\textrm{with}~
c_k[i]=\left(\sum_{l=1}^L\big(x_{k-1}^{(l)}[i]\big)^2\right)^{1/2}
\end{align}
Then formula (\ref{eq-16}) can be rewritten as
\begin{equation}
  X_k=W_kA_k^H(A_kA_k^H+\alpha I)^{-1}Y
\end{equation}
For $\nabla_EJ(x,E)=0$ we can renew (\ref{eq6}) as
\begin{equation}
  E_k=(Y-AX_{k-1})\left(\frac{\sigma^2}{\sigma_2^2}I+X_{k-1}^HX_{k-1}\right)^{-1}  X_{k-1}^H
\end{equation}

\begin{figure}[!ht]
\subfigure[Result recovered by Regularized FOCUSS: weak signal is loss]
{
\label{fig1a}
\begin{minipage}[b]{0.5\textwidth}
  \centering
  \includegraphics[height=6cm]{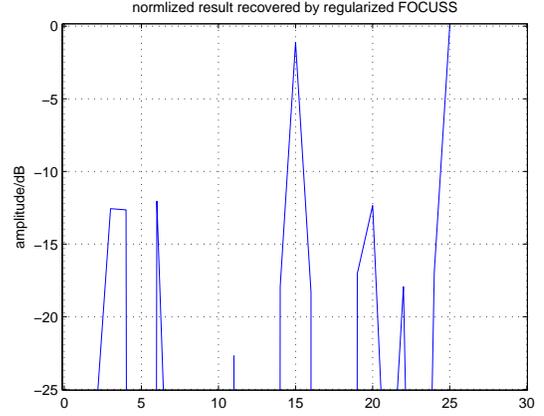}
\end{minipage}}
\subfigure[Result of recovered by TLS-FOCUSS: weak signal is found]
{
\label{fig1b}
\begin{minipage}[b]{0.5\textwidth}
  \centering
  \includegraphics[height=6cm]{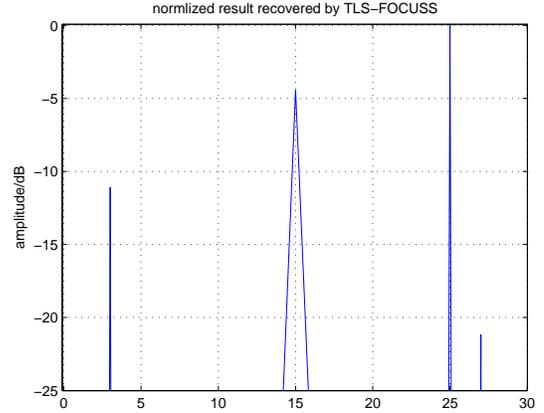}
\end{minipage}}
  \caption{Result of weak signal recovery with $m=20,n=30$}
  \label{fig1}
\end{figure}

Then the \textbf{Algorithm 2} can be modified to fit MMV model as \textbf{Algorithm 3}.

\begin{algorithm}[MMV SD-FOCUSS]
%\caption{MMV SD-FOCUSS}
\mbox{}
\begin{itemize}
\item[] \textbf{Input:} $y$, $\textbf{x}_0$, $E_0$ $A$, $\sigma$, $\sigma_2$, $p$.

\item[1] Set
$W_k=\textrm{diag}\left(\left[c_k[i]^{1-\frac{p}{2}}\right]_{i=1,\cdots,n}\right)$,\\
where $c_k[i]=\left(\sum_{l=1}^L\big(x_{k-1}^{(l)}[i]\big)^2\right)^{1/2}$, $p\in(0,1])$;

\item[2] Calculate
\[E_k=(Y-AX_{k-1})\left[\sigma^2\sigma_2^{-2}I+X_{k-1}^HX_{k-1}\right]^{-1}X_{k-1}^H\] and $A_k=(A+E_{k})W_k$;

\item[3] Calculate $X_k=W_kA_k^H(A_kA_k^H+\alpha I)^{-1}Y$;

\item[4] If $\|X_k-X_{k-1}\|_2^2/\|X_{k-1}\|_2^2<\epsilon$, exit; else goto step 1.
\end{itemize}

\end{algorithm}

\section{Simulation Results}\label{sec5}

The parameters  in this paper are set as: norm-factor $p=0.5$,
convergence threshold $\epsilon=0.01$. In each Monte Carlo simulation,
1000 trials are carried out independently. In each trial, the $m{\times}n$
dictionary $A$ is chosen as Gaussian random matrix, entries of which are independently, identically and normally distributed. In order to analyze the mentioned algorithms, the true sparse solution  has to be known, and it is hard to know in practice problems.

The algorithm in one simulation is considered to be successful if all nonzero-locations of $\textbf{x}$ are
found exactly; otherwise, the algorithm is considered to be failed.

\subsection{Single  Measurement Vector Case}

\begin{figure}[!ht]
\subfigure[success probability of algorithms in finding the support set
  correctly]
  {\label{fig6a}
\begin{minipage}[b]{0.5\textwidth}
  \centering
  \includegraphics[height=6cm]{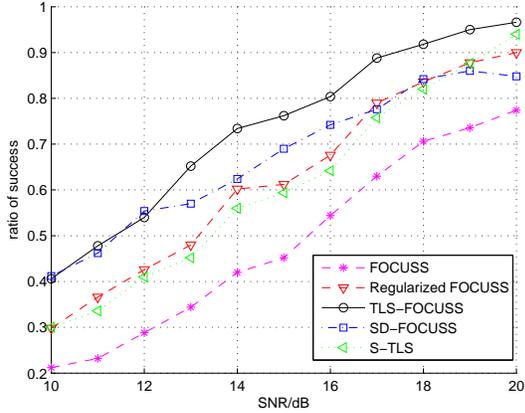}
\end{minipage}}
\subfigure[RMSE of signal amplitude recovery]
  {\label{fig6b}
\begin{minipage}[b]{0.5\textwidth}
  \centering
  \includegraphics[height=6cm]{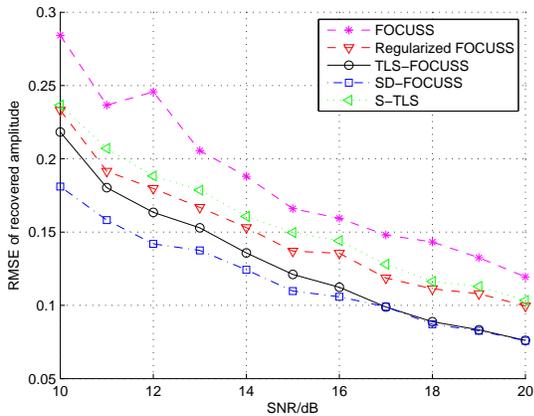}
\end{minipage}}
  \caption{Performance of involved algorithms with $m=20,n=30$} \label{fig6}
\end{figure}

This subsection shows the advantages of recovering ability of
new algorithms from TLS model with numerical simulation. Let $\textbf{x}$ be a
$s$-sparse vector, i.e. $\|\textbf{x}\|_0=s$, and let the average
power of $\textbf{x}$ be normalized, i.e. $\sum_{i} {|x_i|^2}=1$. In each trial, entries of
matrix $[e,E]$ are also independently and identically Gaussian distributed\footnote{if the variances of generalizing
$\textbf{e}$ and $E$ are different, the performance of TLS-FOCUSS
will not change, while the performances of the other algorithms will
be affected.} with mean
zero and variance $\sigma^2$.  Then overall SNR can be represented
as $1/\sigma^2$. The indices of nonzero coordinate set  $T$ are chosen randomly from
a discrete uniform distribution $U(1,N)$ (without repetition).

In following simulations, besides TLS-FOCUSS and SD-FOCUSS,
other algorithms will be involved: standard FOCUSS
\cite{han4}, Regularized FOCUSS \cite{han5}, and S-TLS \cite{han8}.

\begin{figure}[!ht]
\subfigure[percentage success with randiness distribution in sparse entries]
  {\label{success1}
\begin{minipage}[b]{0.5\textwidth}
  \centering
  \includegraphics[height=6cm]{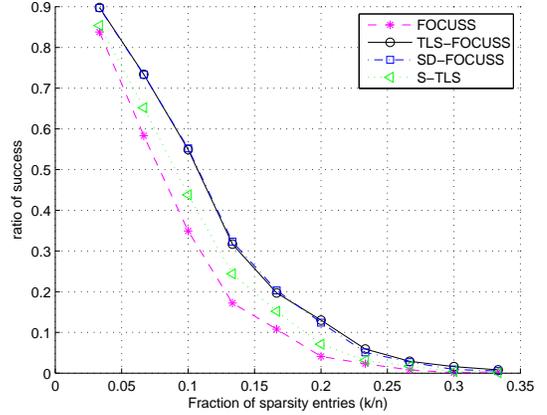}
\end{minipage}}
\subfigure[percentage success with the same amplitude in sparse entries]
  {\label{success2}
\begin{minipage}[b]{0.5\textwidth}
  \centering
  \includegraphics[height=6cm]{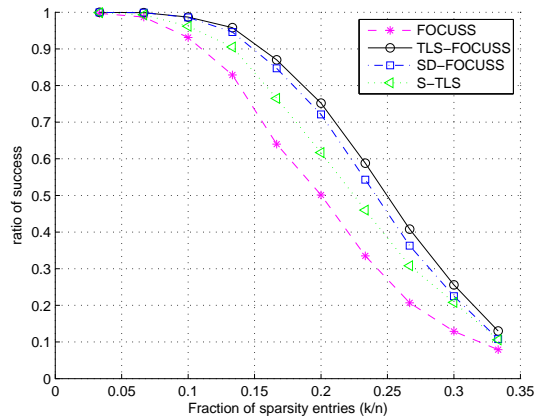}
\end{minipage}}
  \caption{percentage success of involved algorithms with different $k/m$. $m=20,n=30$} \label{success}
\end{figure}

In Fig.~\ref{fig1} and Fig.~\ref{fig6}, the number of rows and
columns of dictionary matrix are set to 20 and 30 respectively. In Fig.~\ref{fig1}, SNR is set to 15 dB,
 $T=[3,15,25]$ and $\textbf{x}_T=[0.4139,-0.9186,-1.4819]^T$. It
can be seen from Fig.~\ref{fig1} that TLS-FOCUSS does much better
than FOCUSS in extracting weak signal when dictionary and
measurement are both corrupted. For TLS-FOCUSS, the position and
amplitude of  signal are both recovered excellently;
 the result of FOCUSS is failed, for weak signal is buried in "False
Peak" brought by perturbation on dictionary and can not be
distinguished correctly.
Fig.~\ref{fig6}(A) shows the statistical results
of percentage success, and Fig.~\ref{fig6}(B) shows the statistical
 root-mean-square error (RMSE) of signal amplitude
recovery when algorithms can find the  nonzero-coordinate $T$
correctly under different SNR scenes. TLS-FOCUSS and SD-FOCUSS are
presented to be more robust from Fig.~\ref{fig6a}, and perform
much better on amplitude recovery from Fig.~\ref{fig6b}.

Fig.~\ref{success} shows the percentage-success curves of algorithms with different $k/m$. In the simulation, $m=20,~n=30,~k=1,2,\cdots,10$, SNR=15dB and entries of $\textbf{x}_T$ are set to obey i.i.d. normal distribution in Fig.~\ref{success1} and 1 in Fig.~\ref{success2}. It can be seen from Fig.~\ref{success} that,     TLS-FOCUSS and SD-FOCUSS perform always better than common algorithms  (FOCUSS) and S-TLS designed to solve fully-perturbed model as $k/m$ changes.

%\begin{figure}[!ht]
%\subfigure[success probability of  algorithms in finding the support set correctly]{
%\label{fig2a}
%\begin{minipage}[b]{0.5\textwidth}
%  \centering
%  \includegraphics[height=5cm]{fig2_a.eps}
%\end{minipage}}
%\subfigure[RMSE of  amplitude recovery for weak, middle and strong signals respectively]
%  {\label{fig2b}
%\begin{minipage}[b]{0.5\textwidth}
%  \centering
%  \includegraphics[height=5cm]{fig2_b.eps}
%\end{minipage}}
%  \caption{Performance of involved algorithms with $m=80,n=512$
%  % , where ($\Box$) obtained by FOCUSS ($m=36$),
%%   ($\bigtriangledown$) obtained by TLS-FOCUSS ($m=36$),
%%   ($\circ$) obtained by TLS-FOCUSS ($m=64$), ($\ast$) obtained by TLS-FOCUSS ($m=64$),
%} \label{fig2}
%\end{figure}

\begin{figure}[!ht]

\begin{minipage}[b]{1.0\linewidth}
  \centering
  \centerline{\epsfig{figure=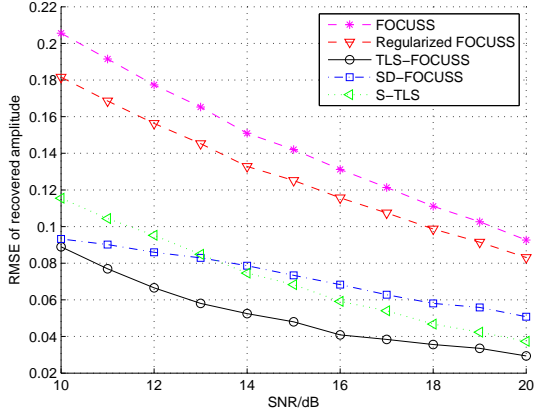,height=6cm}}
  %\centerline{}\medskip
\end{minipage}

  \caption{RMSE of signals recovery in the condition of $m=128$, $n=512$.
} \label{fig3}
\end{figure}

In the simulations of Fig.~\ref{fig3} and Table~\ref{table1}, $m=128$, $n=512$, $s=3$ and $\textbf{x}_T=(1,-1,1)^T/\sqrt{3}$. With smooth curves, Fig.~\ref{fig3} shows that the recovery performance of TLS-FOCUSS in this scenario is much better than the other algorithms; SD-FOCUSS is superior to S-TLS in low SNR, and inferior to S-TLS in high SNR. Table~\ref{table1} shows run-times of mentioned algorithms under the same condition. In order to obtain a measure of the computational complexity, the
average CPU times for each algorithm consumeing is tabulated in
Table~\ref{table1}. It can be seen that, as the same classified algorithms TLS-FOCUSS and SD-FOCUSS are much faster than S-TLS.

By comparison with other algorithms, it can be concluded that TLS-FOCUSS and SD-FOCUSS have the complete advances in percentage succuss, accurate reconstruction and computational speed. And
TLS-FOCUSS has the higher success percentage and more accurate reconstruction than SD-FOCUSS while SD-FOCUSS is  faster than TLS-FOCUSS.

\subsection{MMV Case}
In this simulation we consider the performance of SD-FOCUSS in MMV
case.   $X$ is a  sparse matrix with $L$ columns and only $s$ rows
with nonzero  entries. In each trial, the indices of nonzero rows in
$X$ are chosen randomly from a discrete uniform distribution, and
the amplitudes of the row entries are generalized randomly from a
standard normal distribution; entries of both $E$ and
$[\textbf{e}^{(l)}]_{l=1,\cdots,L}$ are  independently Gaussian
distributed with mean zero and variance $\sigma^2$. The overall SNR
is $1/\sigma^2$. The measurement matrix can expressed as
\begin{equation}
Y=(A+E)X+[\textbf{e}^{(l)}]_{l=1,\cdots,L}\nonumber
\end{equation}
 The relative MSE between the true and
estimate solution is defined as \cite{han16}
\begin{equation}
  \textrm{MSE}=\textrm{E}\left(\frac{\|\hat{X}-X\|_F^2}{\|X\|_F^2}\right)\nonumber
\end{equation}
In following simulations, besides SD-FOCUSS, the other algorithms
will be involved, containing: MMV FOCUSS \cite{han16}, Regularized
MMV FOCUSS \cite{han16}, and MMV OMP \cite{han16}.

\begin{table}[!ht]
\centering
\begin{tabular}{@{\quad}c@{\quad}c@{\quad}c@{\quad}c@{\quad}c@{\quad}c@{\quad}}
 \hline
  SNR &   FOCUSS &   RegFOC &   TLS-FOC &   SD-FOC &   S-TLS\\
  (dB) &   (sec) &   (sec) &   (sec) &   (sec) &   (sec)\\
\hline
10&0.1284& 0.0160& 0.5841&0.3008& 5.1528\\
  11&  0.1298&  0.0182&  0.6530&  0.3513&  5.3670\\
  12&  0.1274&  0.0185&  0.6291&  0.3276&  5.3779\\
  13&  0.1218&  0.0158&  0.5852&  0.3010&  5.2276\\
  14&  0.1215&  0.0156&  0.6001&  0.2964&  5.2652\\
  15&  0.1211&  0.0156&  0.5863&  0.2959&  5.3563\\
  16&  0.1202&  0.0155&  0.5858&  0.2961&  5.3963\\
  17&  0.1213&  0.0156&  0.5867&  0.2958&  5.3639\\
  18&  0.1211&  0.0155&  0.5880&  0.2966&  5.3069\\
  19&  0.1213&  0.0163&  0.6104&  0.3121&  5.2285\\
  20&  0.1221&  0.0156&  0.5921&  0.2965&  5.1877\\
\hline
\end{tabular}

\caption{Run-time of algorithms with $m=128,n=512$. The simulations are done in Matlab 7.8 on a Core 2, 3.0-GHz, 2-GByte RAM PC}\label{table1}
\end{table}

\begin{figure}[!ht]
\centering
%\begin{minipage}[t]{0.5\linewidth}
  \centering
  \includegraphics[height=6cm]{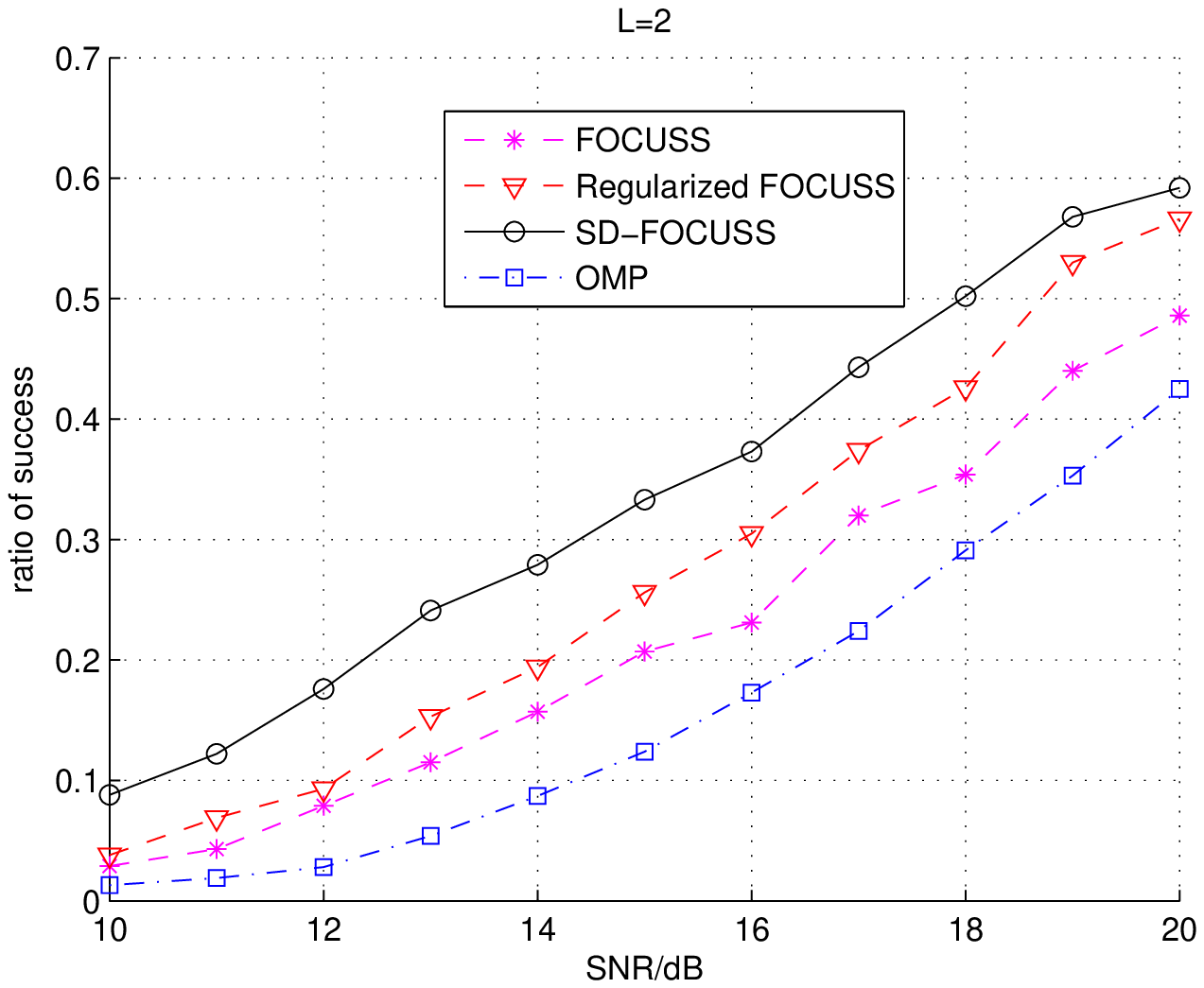}\hspace{4em}
  \includegraphics[height=6cm]{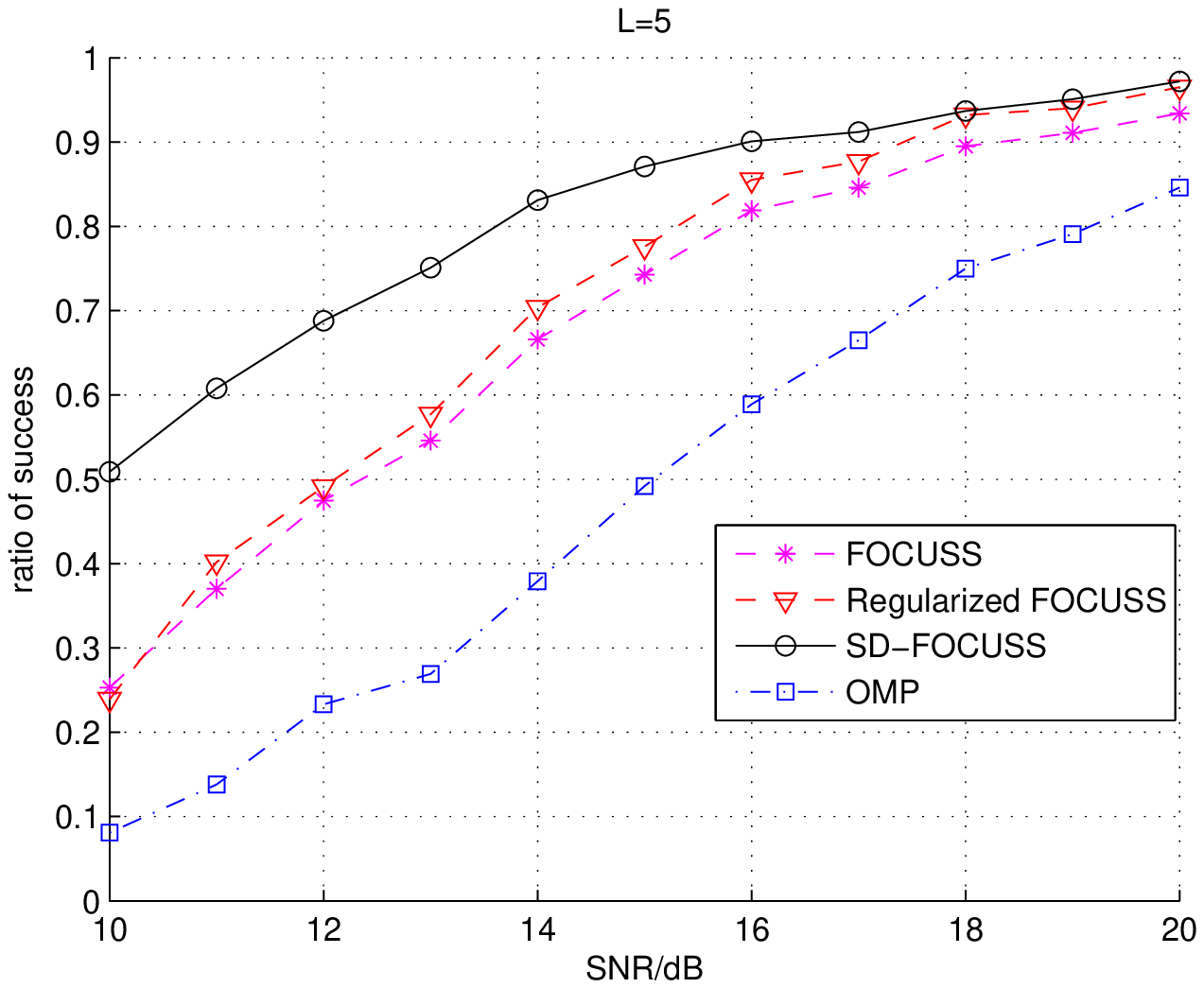}\hspace{4em}
  \includegraphics[height=6cm]{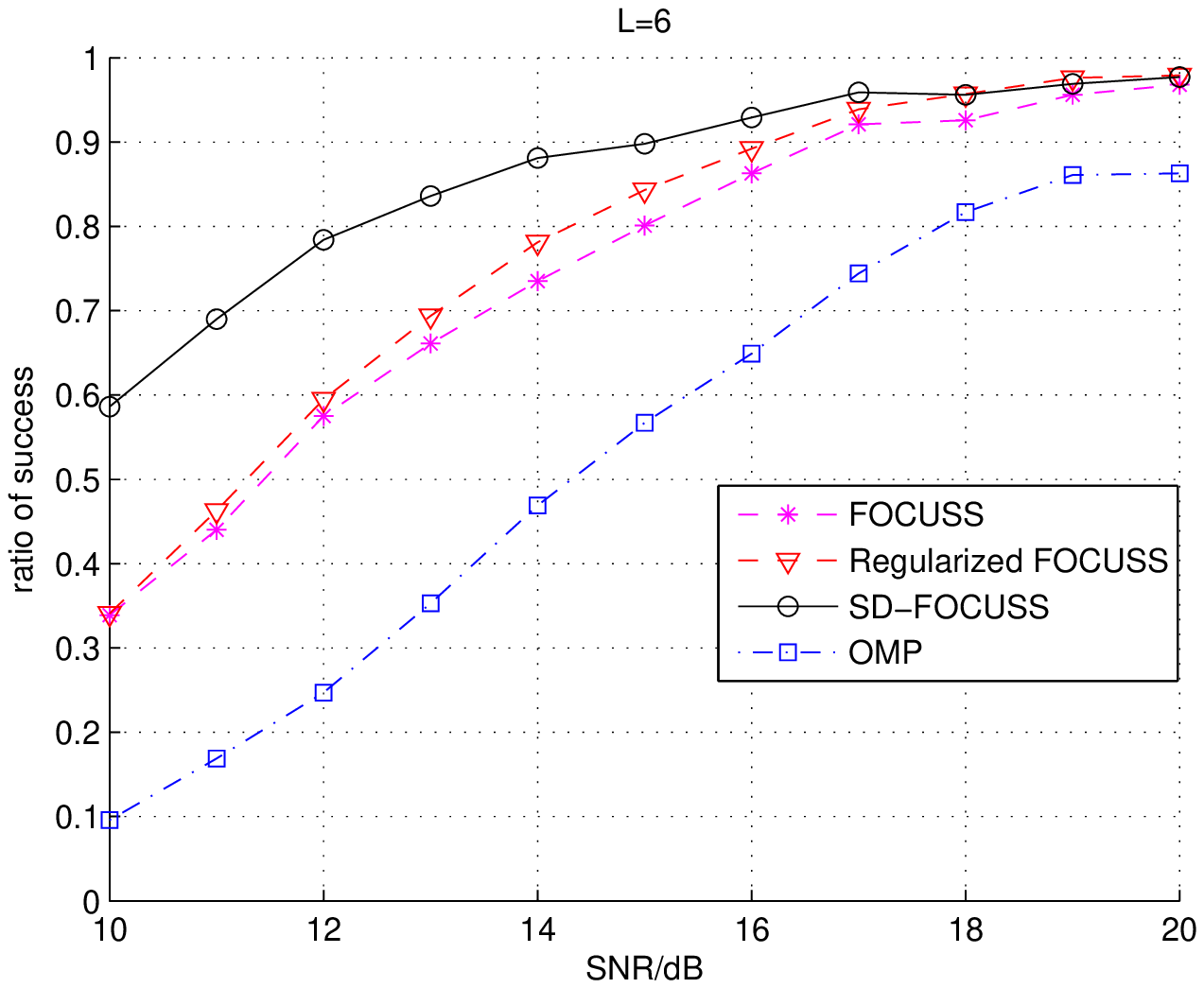}\hspace{4em}
%\end{minipage}
  \caption{Success probability of algorithms obtaining
  all $s$ nonzero rows in MMV case, with $m=20,~n=30,~\textrm{s}=7$ and
  Number of observation vectors is set to $L=2,5,6$.
} \label{fig4a}
\end{figure}

\begin{figure}[!ht]
  {
%\begin{minipage}[t]{0.5\linewidth}
  \centering
  \includegraphics[height=6cm]{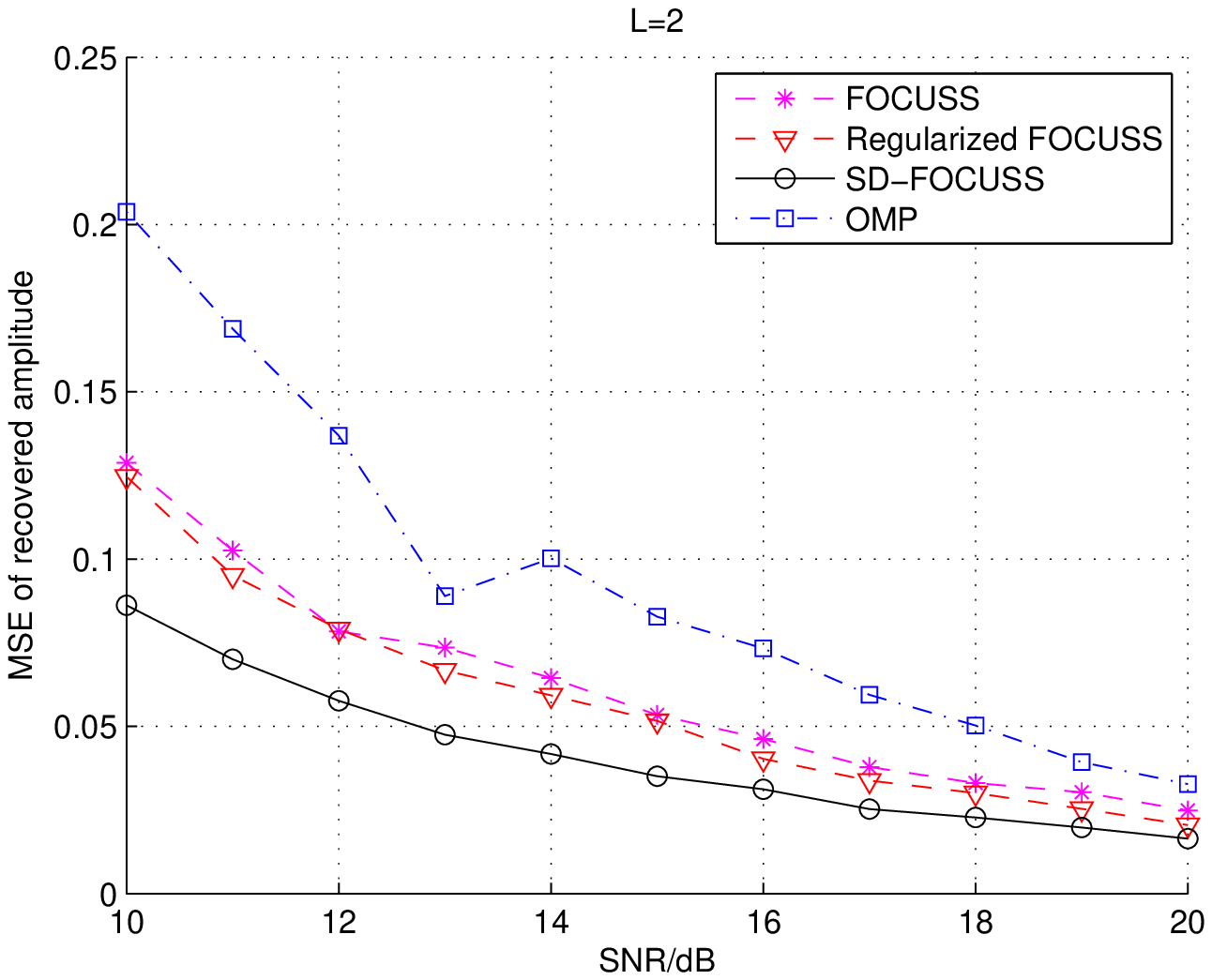}\hspace{4em}
  \includegraphics[height=6cm]{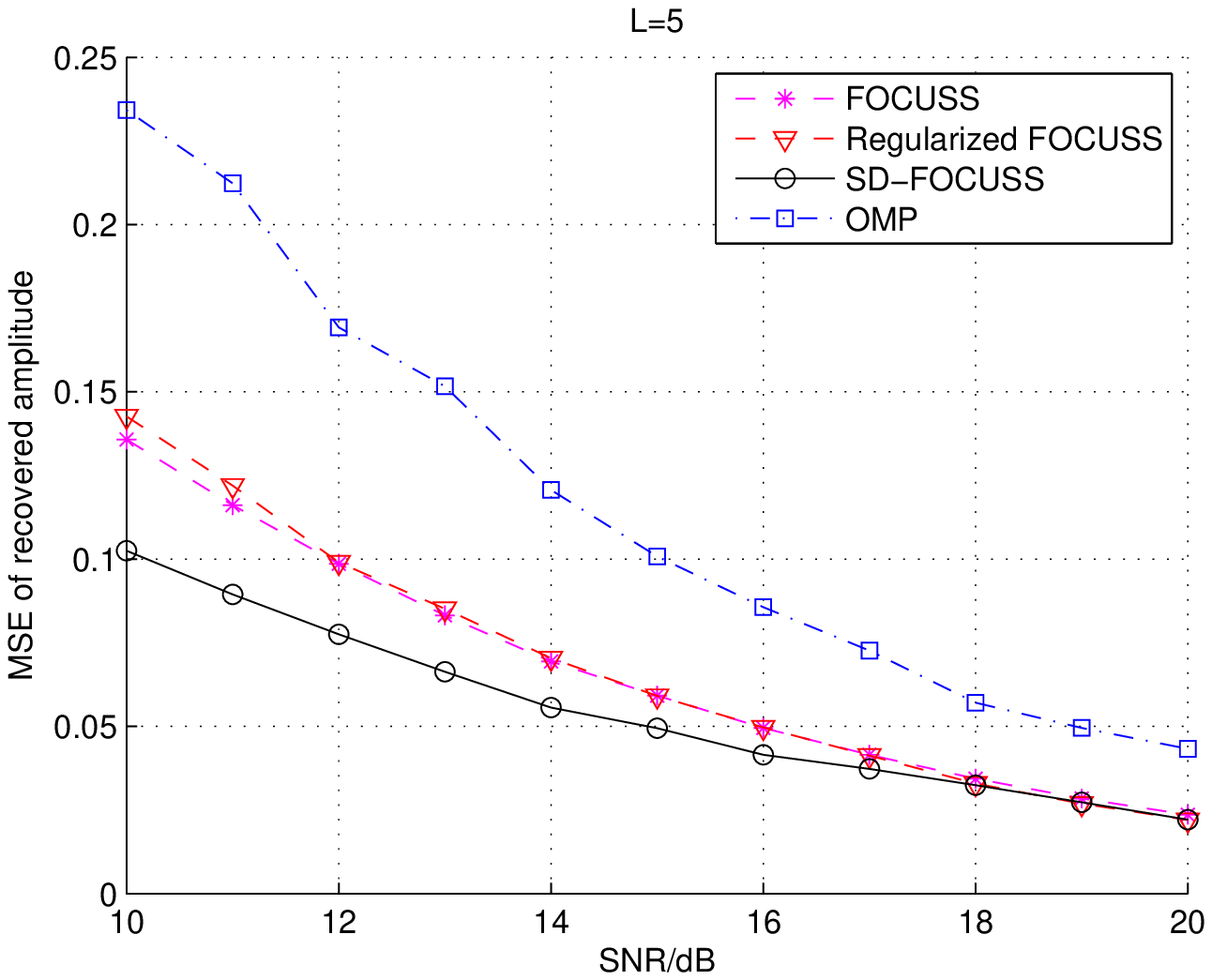}\hspace{4em}
  \includegraphics[height=6cm]{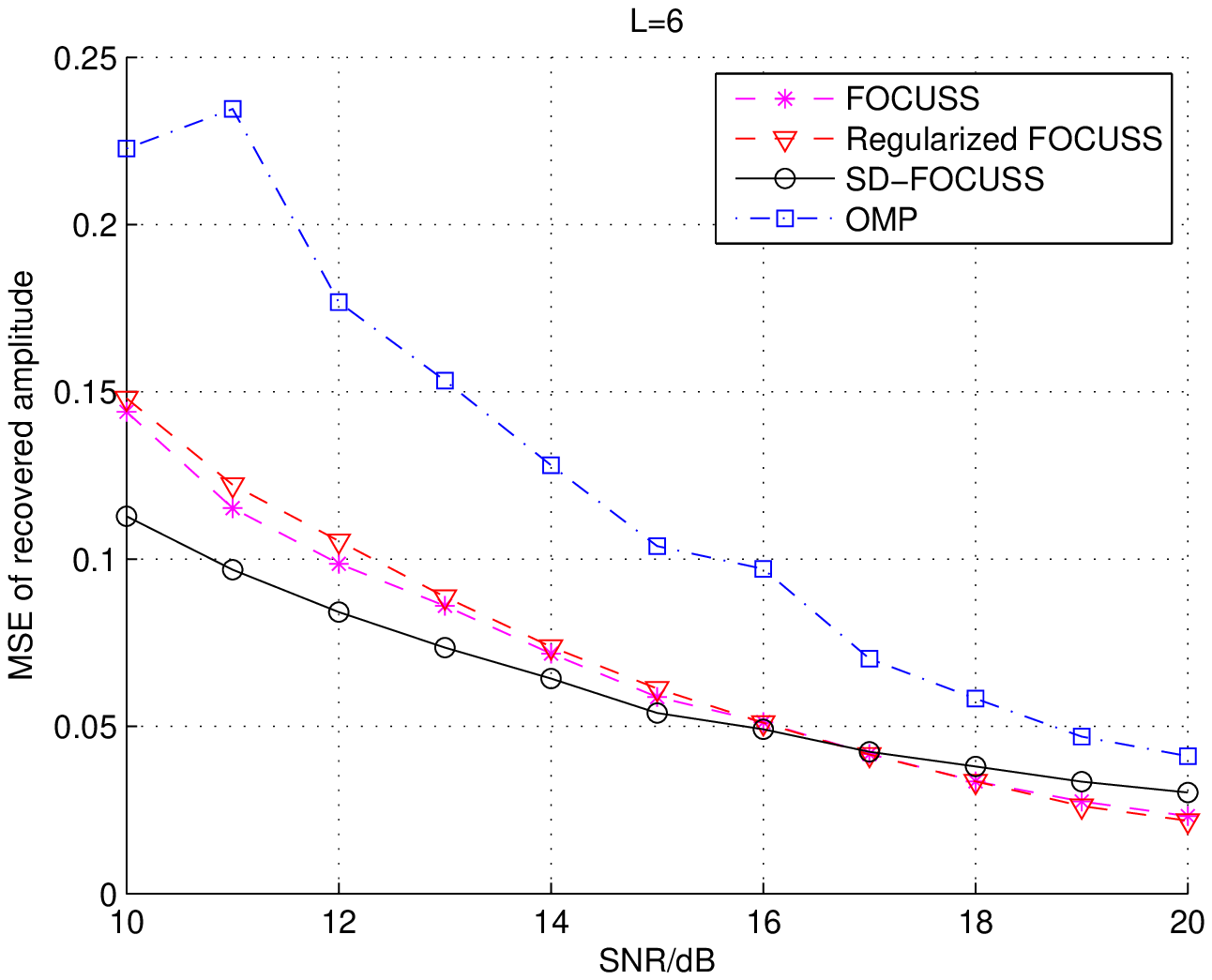}\hspace{4em}
%\end{minipage}
}
  \caption{Relative MSE of amplitude recovery in MMV case, with $m=20,~n=30,~\textrm{s}=7$ and
  Number of observation vectors is set to $L=2,5,6$.
} \label{fig4b}
\end{figure}

The number of rows and columns of dictionary $A$ are set to 20 and
30 respectively, and let $s=7$. Two quantities are varied in this
experiment: SNR and $L$. Fig.~\ref{fig4a} and Fig.~\ref{fig4b} show
success-probability curves and MSE curves respectively when
$L=2,5,6$. It can be found that as $L$ becomes larger, success
numbers become larger; however, MSE curves seem to be unchanged for
it is related with perturbation and unrelated with $L$. MMV
SD-FOCUSS performs better than other algorithms.

\section{Conclusion}\label{sec6}

In this paper, through extending FOCUSS algorithms, we have proposed
two new algorithms, TLS-FOCUSS and SD-FOCUSS, to recover the sparse
vector from an underdetermined system when the measurements and
dictionary matrix are both perturbed. The convergence of algorithms
was considered. Then we applied SD-FOCUSS in MMV model with a row-sparsity structure.  The simulations showed
our approaches performed  better than   other present algorithms  in computational complexity, percentage success  and RMSE of signal
amplitude recovery. The
benefits of TLS-FOCUSS and SD-FOCUSS make them be  good candidates of
sparse recovery algorithms for more practical applications.

%\appendices
%\section{23132123}
%dfgdfg

\bibliographystyle{IEEEbib}
\bibliography{strings}

%\end{CJK*}
\end{document}